\documentclass[showpacs,aps,prl,twocolumn,superscriptaddress]{revtex4-1}
\usepackage{graphicx} 
\usepackage{dcolumn}
\usepackage{bm}
\usepackage{array}
\usepackage{amssymb,amsmath}
\usepackage{epstopdf}
\usepackage{color}
\usepackage{mathrsfs}
\def\etal{$\it{et~al.}$}
\newcolumntype{C}[1]{>{\centering\arraybackslash}p{#1}}

\begin{document}
\title{Electron-Hole Asymmetry of Surface States in Topological Insulator Sb$_2$Te$_3$ Thin Films Revealed by Magneto-Infrared Spectroscopy}
\author{Y. Jiang}
\affiliation{National High Magnetic Field Laboratory, Tallahassee, Florida 32310, USA}
\email{yjiang@magnet.fsu.edu, liangwu@sas.upenn.edu}
\author{M. M. Asmar}
\affiliation{Department of Physics and Astronomy, The University of Alabama, Tuscaloosa, Alabama 35487, USA}
\author{X. Y. Han}
\affiliation{Department of Physics and Astronomy, University of Pennsylvania, Philadelphia, Pennsylvania 19104, USA}
\author{M. Ozerov}
\affiliation{National High Magnetic Field Laboratory, Tallahassee, Florida 32310, USA}
\author{D. Smirnov}
\affiliation{National High Magnetic Field Laboratory, Tallahassee, Florida 32310, USA}
\author{M. Salehi}
\affiliation{Department of Material Science and Engineering, Rutgers, The State University of New Jersey,Piscataway, New Jersey 08854, USA}
\author{S. Oh}
\affiliation{Department of Physics and Astronomy, Rutgers, The State University of New Jersey,
Piscataway, New Jersey 08854, USA}
\author{Z. Jiang}
\affiliation{School of Physics, Georgia Institute of Technology, Atlanta, Georgia 30332, USA}
\author{W.-K. Tse}
\affiliation{Department of Physics and Astronomy, The University of Alabama, Tuscaloosa, Alabama 35487, USA}
\author{L. Wu}
\affiliation{Department of Physics and Astronomy, University of Pennsylvania, Philadelphia, Pennsylvania 19104, USA}
\date{\today}

\begin{abstract}
  When surface states (SSs) form in topological insulators (TIs), they inherit the properties of bulk bands, including the electron-hole (e-h) asymmetry but with much more profound impacts. Here, via combining magneto-infrared spectroscopy with theoretical analysis, we show that e-h asymmetry significantly modifies the SS electronic structures when interplaying with the quantum confinement effect. Compared to the case without e-h asymmetry, the SSs now bear not only a band asymmetry as that in the bulk but also a shift of the Dirac point relative to the bulk bands and a reduction of the hybridization gap up to 70\%. Our results signify the importance of e-h asymmetry in band engineering of TIs in the thin film limit.
\end{abstract}

\maketitle
Three dimensional (3D) topological insulators (TIs) have gained wide interests in the unique properties of their surface states (SSs). These states result from the difference in topological phase between its bulk and surrounding environment \cite{TI_review_0,TI_review_1}. Ideally, these SSs can be effectively described by the Dirac equations where two linear bands cross at the Dirac point (DP) with  spins locked to their momenta. Such spin helicities greatly suppress back scattering to non-magnetic disorders in transport, rendering TIs  great platforms for future device applications such as low dissipation electronics and quantum computing \cite{spintronics,QComp}. 

Successful device applications of SSs require detailed knowledge of their band structures in the thin film limit. It is now well understood in both theory and experiment that, when the film is reduced to a thickness comparable with the size of SS wavefunctions, a band gap naturally appears due to the hybridization between top and bottom SSs \cite{Gap_XY,Gap_BiSe,Gap_SbTe,FD,Osci_1,Osci_2,eh_Theory1,Theory2,Theory3}. However, other aspects of SSs in the thin film limit are mostly investigated by theories and largely unexplored by experiments, \cite{FD,Osci_1,Osci_2,eh_Theory1,Theory2,Theory3} such as the effects on SSs of electron-hole (e-h) asymmetry and its interplay with the quantum confinement effect \cite{Theory2}. Addressing these questions experimentally is important and provides insights into band engineerings for future TI devices in both equilibrium and non-equilibrium states \cite{Frank1,Frank2}.

Magneto-infrared (IR) spectroscopy is a powerful tool for precise electronic structure characterization \cite{YJ_ZT5,YJ_Gr}. Due to high carrier densities in previous TI systems \cite{Bulk1,Bulk2}, magneto-IR studies did not succeed in observing clear signatures of Landau level (LL) transition between the Dirac SSs \cite{Orlita,ShaoYM,IR1,IR2}. An effective way to suppress the excess carrier densities is to grow ternary or quartnary alloys \cite{Quart1,Quart2,Tern1}. Alternatively,  a more precise and repeatable control over carrier densities in the thin film limit is to use interfacial engineering and compensation doping in molecular beam epitaxy (MBE) \cite{MBE1,Oh_AM}. In this approach, a virtual substrate is grown to match the lattice constant of the target material, thus reducing the defect density at the interface. The carrier density can be further suppressed and even tuned between different carrier types through compensation doping. The TI thin films grown by this MBE approach bear an order of magnitude lower density than alloy TIs, permitting the observations of intriguing topological quantum effects such as the quantum Hall effect and quantized Faraday/Kerr rotations \cite{MBE1,Liang,WKT1,SR1,SR2}.

In this work, we performed magneto-IR spectroscopy in two MBE grown Sb$_2$Te$_3$ thin films with different thicknesses. Owing to the ultralow carrier densities in these samples, we clearly observed Landau level transitions related to the topological SSs. By combining our experiments with a self-consistent modeling approach, we show that the e-h asymmetry and its interplay with the quantum confinement effect have significant impacts on all aspects of the SSs: not only the hybridization gap but also the band dispersion, band asymmetry and their relative energy positions with the bulks. Our results signify the importance of e-h asymmetry and also provide a new twist in SS engineering in low dimensions.

\begin{figure*}[t!]
\includegraphics[width=16cm] {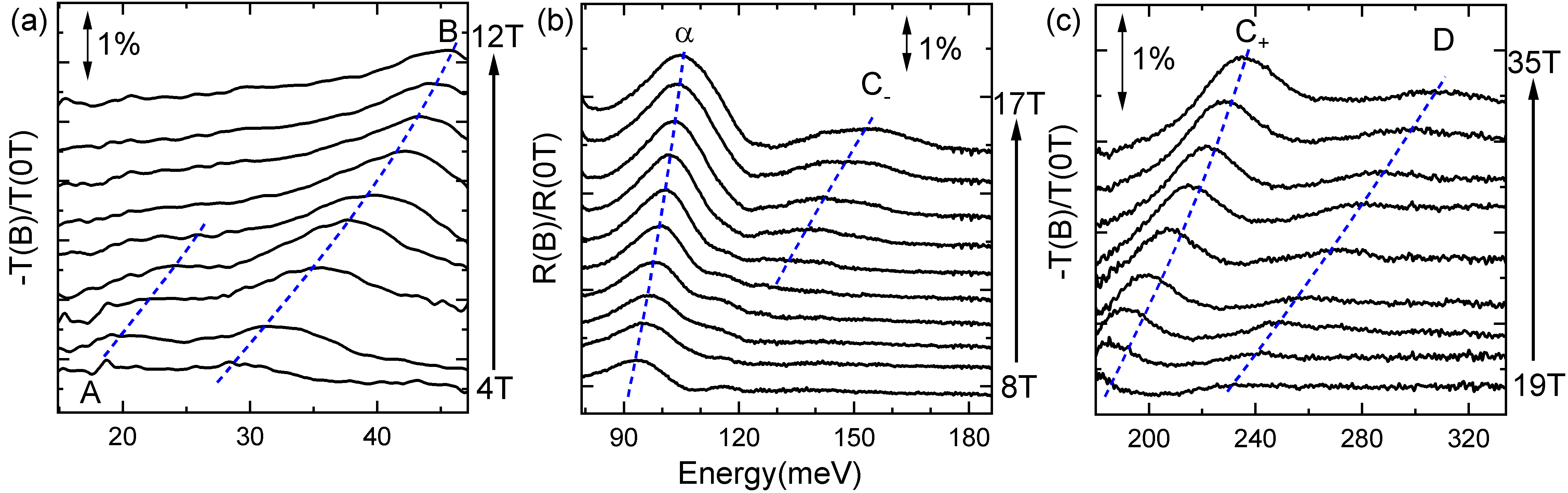}
\caption{(color online) Normalized magneto-IR spectra of the 10-QL sample for (a)(c) transmission and (b) reflection measurements performed outside and inside the substrate Restrahlen band, respectively. All the modes are labeled with a Latin or Greek letter. The blue dash lines are guides to the eye showing the mode evolution in magnetic fields. All spectra are shifted for clarity.}
\end{figure*}

We grow by MBE two Sb$_2$Te$_3$ thin films of 10 and 8 quintuple layers (QLs) on a sapphire (Al$_2$O$_3$) substrates. The thin films are sandwiched between a buffer layer of In$_2$Se$_3$/(Sb$_{0.65}$In$_{0.35}$)$_2$Te$_3$ to reduce the interface defects and a capping layer of (Sb$_{0.65}$In$_{0.35}$)$_2$Te$_3$ to maintain a symmetric environment between their bottom and top surfaces. From transport measurements, these samples show record low carrier densities with $n_e= 2.2 \times 10^{11}$ cm$^{-2}$ (n-type) and $n_h=3.1 \times 10^{11}$ cm$^{-2}$ (p-type) for the 10-QL and 8-QL samples, respectively, which are even lower than those in Bi$_2$Se$_3$ MBE samples combined with charge transfer doping technique\cite{MBE1,Liang}. Further information on sample characterizations and growth can be found in Ref. \onlinecite{Oh_AM} and also Supporting Information. For magneto-IR measurements, 
since the sapphire substrate has a strong Reststrahlen band over a wide energy range (50-180 meV), we combine both transmission and reflection measurements to cover a complete spectra from far- to mid-IR. The samples are illuminated with IR light from a globar source guided via vacuummed lightpipes. The transmitted/reflected lights are collected afterward with a Si-composited bolometer detector. Transmission measurements are performed in a 35 T resistive magnet, while reflection measurements are conducted in a 17.5 T superconducting magnet. All the measurements are performed in the Faraday configuration with a magnetic field (B) perpendicular to the films at liquid helium temperature 4.2 K.

Our theoretical modeling of the SSs employs a self-consistent approach by considering the bulk Hamiltonian with proper boundary conditions, which allows us to properly address the relative position between the bulk bands and the SSs. We start from the well-known effective TI Hamiltonian for bulk ($H_B$) up to $k^2$ terms:\cite{BHZ_TImodel,BHZ_Buttner,Orlita,Dmitry}
\begin{align*}
    H_B(\mathbf{k})=&\epsilon(\mathbf{k})\tau_0\sigma_0+M(\mathbf{k})\tau_z\sigma_0+B_0k_z\tau_y\sigma_0\\
    &+A_0(k_y\tau_x\sigma_x-k_x\tau_y\sigma_y),
\end{align*}
where $\epsilon(\mathbf{k})=C_0+C_1k_z^2+C_2( k_x^2+k_y^2)$, $M(\mathbf{k})=M_0+M_1k_z^2+M_2(k_x^2+k_y^2)$, and $z$ is defined as the perpendicular direction to the thin film. Here, $\mathbf{\tau}$ and $\mathbf{\sigma}$ are the Pauli matrices for orbitals and spins, respectively, and both $\tau_0$ and $\sigma_0$ are the identity matrix. The bulk electronic band structure is characterized by a band gap $2M_0$, linear band components $A_0$, $B_0$, and parabolic band components $M_1$, $M_2$. The $\epsilon(\mathbf{k})$ term describing the e-h asymmetry also plays an important role as will be shown next. The resultant SSs can be captured by replacing the $k_z$ term with the differential operator $-i\partial/\partial{z}$ and applying the open boundary conditions at the two end surfaces along the $z$-axis. Meanwhile, the $k_x$ and $k_y$ still remain good quantum numbers due to their translational invariance in plane. We solve the boundary value problem numerically to get zero field band structures \cite{FD,YJ_InAs}. 

It is instructive to see how the SSs are affected by the e-h asymmetry term $\epsilon(\mathbf{k})$ from the bulk. One can arrive at an effective Hamtiltonian for SSs $H_{ss}$ from $H_B$ if the film thickness $L$ is in the thick and ultrathin limits \cite{Theory2,eh_Theory1}:
 \[
     H_{ss}= \left\{
         \begin{array}{l}
        (C_0-\frac{M_0}{M_1}C_1)\sigma_0+(C_2-\frac{M_2}{M_1}C_1)k^2\sigma_0\\+A_0\sqrt{1-\frac{C_1^2}{M_1^2}} (\mathbf{\sigma} \times \mathbf{k})_z, \qquad\qquad \text{if  } L \rightarrow +\infty,\\
         C_1 \frac{\pi^2}{L^2}\sigma_0+C_2 k^2\sigma_0+A_0 (\mathbf{\sigma} \times \mathbf{k})_z\\- (M_1\frac{\pi^2}{L^2}+M_2k^2) \tau_z\sigma_z, \qquad\qquad \text{if  } L \rightarrow 0.
        \end{array}
        \right.
  \]
Without the bulk e-h asymmetry (i.e., $C_0,C_1,C_2=0$), the SSs are described by the linear term ($\propto (\mathbf{\sigma} \times \mathbf{k})_z$) and a gap $\left| 2M_1 \pi^2/L^2 \right|$ in the ultrathin limit as massless/massive Dirac fermions with symmetric conduction and valence SSs. With the bulk e-h asymmetry, the constant terms ($\propto \sigma_0$) produce an energy shift in the DP, which is important to the relative positions with the bulk bands; the $k^2 \sigma_0$ terms lead to a band asymmetry between the conduction and valence SS dispersion, that is, a positive (negative) band asymmetry for a faster electron (hole) band. In addition, it is easy to identify that these terms are dependent on the film thickness by comparing them in the two limits. We shall return to this point with a more detailed discussion later.

In the presence of external magnetic field along the $z$ direction, we include a Zeeman Hamiltonian $H_z=\mu_{0}B [(g_1-g_2)\tau_{z}\sigma_{z}+(g_1+g_2)\tau_0\sigma_{z}]/4$ to $H_B$, where $g_1$ and $g_2$ are the effective $g$-factors, and $\mu_0$ is the Bohr magneton. Then, we use the standard Peierls substitution and the wavefunction ansatz in Ref. \onlinecite{BHZ_TImodel} to rewrite the $k_x$ and $k_y$ in terms of LL index $n$ and magnetic field $B$. Now, the LLs can be obtained by similar procedures in the zero field calculation. In this model, the selection rules are $\Delta n=\pm1$, where the LL index changes by 1 between the initial and final states \cite{SR1,SR2}. In principle, there is an additional selection rule related to a quantum number differentiating the LLs within the same $n$ \cite{SR3}. However, its difference is not present until very high magnetic fields and is not discernible in our experiments due to the broad linewidth. Therefore, we omit its labeling in our assignment. In our calculation, we find the following parameters give the best description of our experiments: $A_0=3.4\text{ eV\r{A}}, B_0=1.8\text{ eV\r{A}}, M_0=-0.22\text{ eV}, M_1=22.64\text{ eV\r{A}}^2, M_2=48.51\text{ eV\r{A}}^2, C_0=0.001\text{ eV}, C_1=-12.39\text{ eV\r{A}}^2, C_2=-10.78 \text{ eV\r{A}}^2, g_1=-14.45$, and $g_2=14.32$. It is important to emphasize that our fitting procedure starts with the values from the first-principles calculation \cite{BHZ_TImodel}, and by only adjusting the $B_0$ and $M_1$ parameters, we can find a common set of parameters for both samples with different thicknesses as their only variable in the modeling.

\begin{figure}[t]
\includegraphics[width=8.5cm] {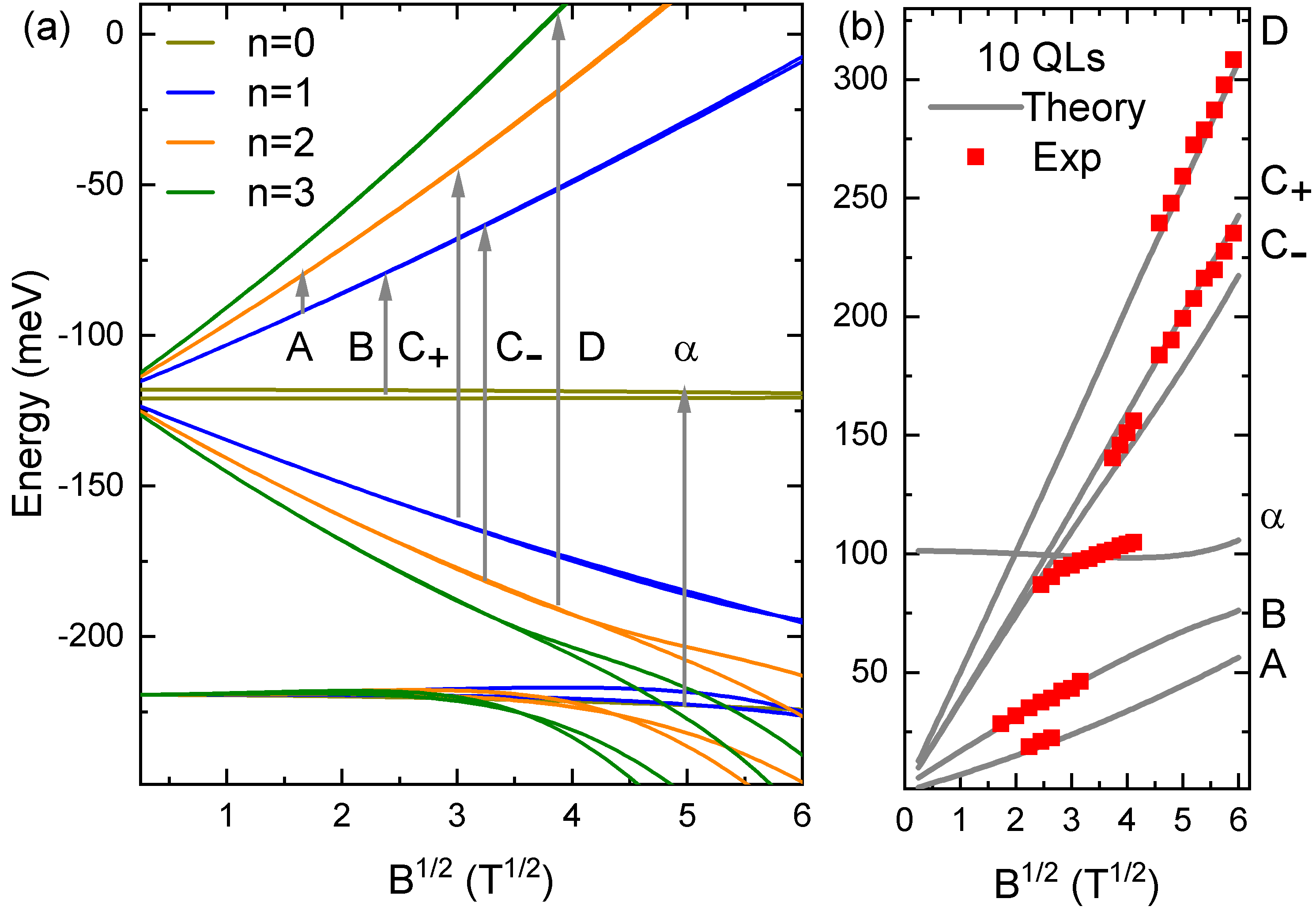}
\caption{(color online) (a) Calculated LLs for the 10-QL Sb$_2$Te$_3$ sample using the parameters listed in the main text. The LL index $n$ is indicated by different colors. (b) Theoretical fits to the observed LL transitions. The corresponding transitions are labeled by the arrows in (a). For simplicity, we only consider the transitions between $\Delta n=1$ and $\Delta n=-1$  as discussed in the main text.}
\end{figure}

Figure 1 shows the normalized magneto-IR spectra of the 10-QL Sb$_2$Te$_3$ thin film in different magnetic fields and energy regions. The LL transitions emerge as spectral peaks which blue shift with increasing magnetic fields. Their evolutions in magnetic fields are indicated by the blue dash lines. By visual inspection, the $\alpha$ mode shows a much weaker magnetic field dependence than those of other modes, already indicating its distinct origin. These modes, except for the A mode, can also be identified in the 8-QL sample but at slightly different energy positions (see Supporting Information).

\begin{figure}[t]
\includegraphics[width=8.5cm] {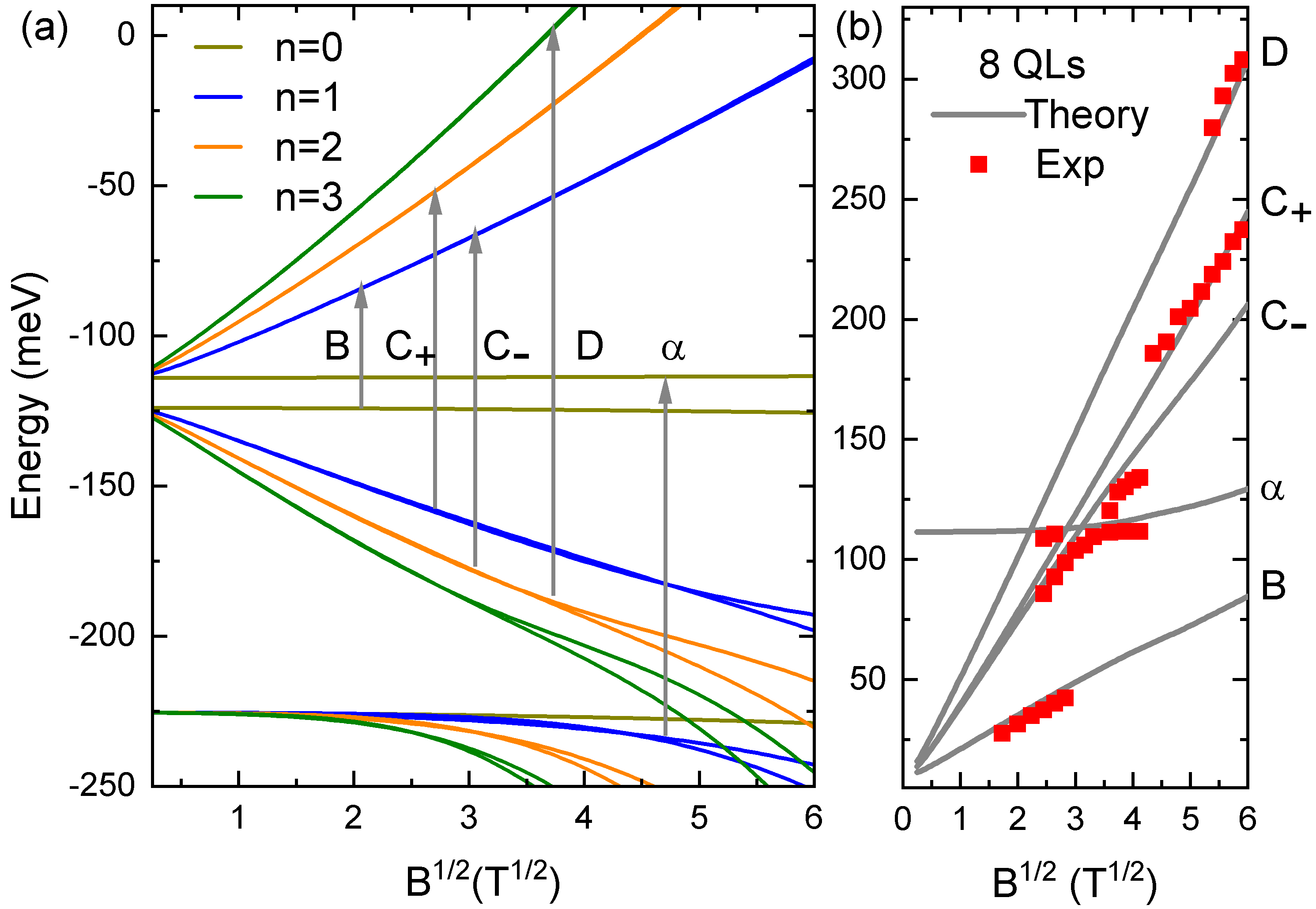}
\caption{(color online) Same as Fig. 2 but with the 8-QL sample.}
\end{figure}

 Figures 2 and 3 show the calculated LLs and compare their corresponding transition energies to the experiment data. We find that all the modes can be separated into two groups as shown in Figs. 2(b) and 3(b). The A, B, C$_-$, C$_+$, and D modes form one group since they all exhibit a clear linear dependence in $\sqrt{B}$ and collapse into the same intercept close to zero energy at $B=0$ T. Their slopes in the $E$ vs $\sqrt{B}$ plot closely match the ratio  $1:1+\sqrt{2}:\sqrt{2}+\sqrt{3}$, a typical $\sqrt{n}$ dependence between different interband transitions of a Dirac fermion \cite{YJ_ZT5}. Since the SSs in these TI systems are the only known ones to exhibit a Dirac type dispersion, we can assign these modes to the LL transitions between the SSs as labeled in Figs. 2(a) and 3(a). Specifically, for the 10-QL sample, the lowest energy mode A is assigned to a cyclotron resonance between $n=1$ and $n=2$ LLs. The A mode disappears above 7 T when the carriers in the $n=1$ LL are completely emptied out. From here, we can estimate the carrier density to be $|n_e|=1.7 \times 10^{11} /$cm$^2$, in agreement with the transport measurement. For both samples, the B, C$_-$, C$_+$, and D modes are assigned to interband LL transitions between the valence and conduction SS. To further attest the validity in the assignment, we can extract a Fermi velocity of $v_F=4.5 \times 10^{5}$ m/s from the slope, consistent with a previous STM measurement \cite{Gap_SbTe}. 
 
A closer inspection of the C$_-$ and C$_+$ modes finds that their energies are not well aligned on the same line. In fact, they can be taken as a splitting between the $\left|\Delta n \right| =1$ transitions as a result of the e-h asymmetry in the SSs. As the correction from the $k^2 \sigma_0$ term breaks the band symmetry between the conduction and valence SSs, the degeneracy between $\Delta n=-1$ (C$_-$ mode) and $\Delta n=1$ (C$_+$ mode) LL transitions are lifted, leading to the observed splitting\cite{YJ_ZT5,YJ_Gr}. In addition, the splitting between the C$_-$ and C$_+$ mode is more prominent in the 8-QL sample, suggesting that the band asymmetry is enhanced as the film thickness is reduced. In principle, the D mode should also exhibit a similar splitting. However, its broad linewidth and weak intensity prevent us from discerning it.

On the other hand, the $\alpha$ mode bears saliently different features from the other modes. Even though it follows a $\sqrt{B}$ dependence, the slope of the $\alpha$ mode does not fit into the SS transition sequence. More importantly, it yields an 80 meV of zero field intercept by linear extrapolation. This number is much larger than zero but smaller than the 300 meV Sb$_2$Te$_3$ bulk gap \cite{Gap_SbTe}, strongly suggesting it as a transition between the SS and the bulk band. In addition, the fact that the intercept is less than half of the bulk gap indicates a significant shift of the DP due to e-h asymmetry, which otherwise should be lying at the center between the conduction and valence bulk bands\cite{Theory3,Theory4}. As is obvious, the relative position between the DP and the bulk bands is essential in understanding the intercept in the surface-to-bulk transition. This is only possible with our self-consistent approach but not in models considering only SSs \cite{Theory2,Theory3,SR3}. Finally, when the thickness reduces from 10 QLs to 8 QLs, the intercept of $\alpha$ mode increases. We will see immediately that this shift is attributed to the hybridization gap enhancement. 

\begin{figure}[t]
\includegraphics[width=8.5cm] {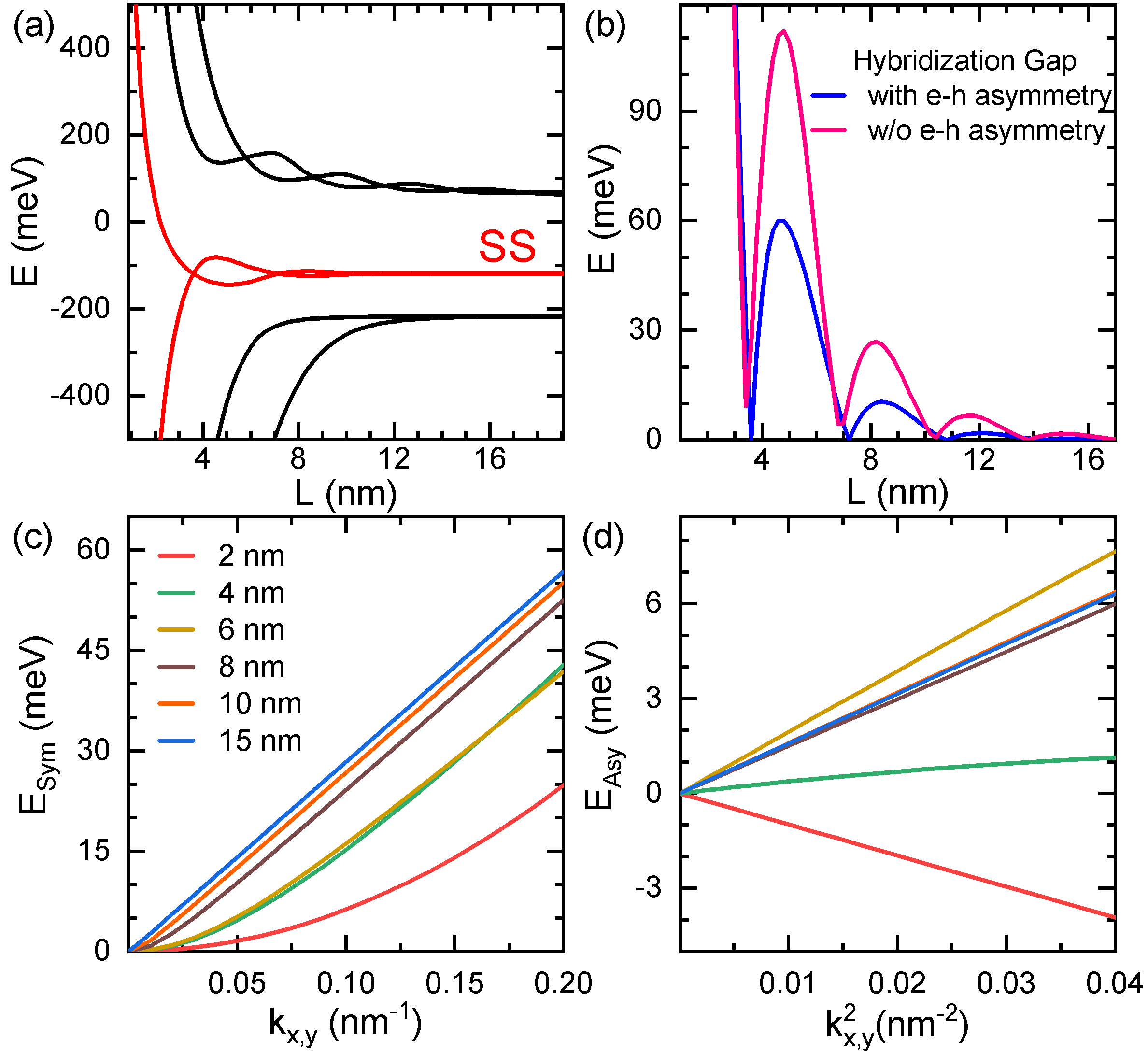}
\caption{(color online) Calculated thickness dependence for (a) the energies of the low-lying bulk states (black) and SSs (red) at $\Gamma$ point and (b) the hybridization gap with (blue) and without (pink) e-h asymmetry. (c) The symmetric ($E_{Sym}$) and (d) asymmetric ($E_{Asy}$) components in the SS dispersions at different thicknesses. In (c) and (d), all the curves are shifted for clear comparisons. The parameters used in the calculation are the same as those extracted from experiments.}
\end{figure}

Further insights into the SSs dependence on e-h asymmetry, thickness and their interplay can be obtained by modeling their behaviors based on our realistic parameters. Our SSs analysis includes not only the hybridization gap but also their dispersions, the latter of which are mostly neglected in previous work \cite{FD,Osci_1,Osci_2,Theory3}. Figure 4(a) shows the thickness dependence of the energy positions for the low-lying bulk bands and the SSs at $\Gamma$ point. Due to the e-h asymmetry, the SSs are lying closer to the valence bulk band before they diverge below $\sim$ 4 nm. Their relative positions remain mostly unchanged down to 7 nm, which explains why the shift in $\alpha$ mode with reducing thickness is interpreted as a hybridization gap enhancement. Meanwhile, the SSs gradually open a hybridization gap that is noticeable at the largest of $\sim$ 14 QLs as shown in Fig. 4(b), indicating the system experiences a dimensional crossover from a 3D TI into a 2D insulator. The oscillatory behaviors in Fig. 4(a)(b) are related to alternating topological phase transitions between a 2D quantum spin Hall and a topologically trivial insulator state at different thicknesses \cite{FD,Osci_1,Osci_2,Theory2,Theory3}. By comparing the thickness dependence of the gap with and without e-h asymmetry in Fig. 4(b), we find that the thicknesses for gap minima in the oscillations remain about the same, but the gap size can be drastically reduced by up to 70\% with e-h asymmetry. Figure 4(c)(d) calculate the symmetric ($E_{Sym}$) and asymmetric ($E_{Asy}$) components of the SS dispersion by $E_{Sym}=(E_{\text{CSS}}-E_{\text{VSS}})/2$, and $E_{Asy}=(E_{\text{CSS}}+E_{\text{VSS}})/2$.
Here, $E_{\text{CSS}}$ and $E_{\text{VSS}}$ are the dispersions in the conduction and valence SS, respectively (see also Supporting Information). The symmetrized component represents the dominant dispersion of the SS, where linear dispersion is gradually replaced by parabolic dispersion as the thickness is reduced. This is because the hybridization gap is so large in the ultrathin limit that SSs can no longer be described by the Dirac fermion picture. On the other hand, the anti-symmetrized component represents the band asymmetry in the dispersion as they show a clear $k^2$ dispersion. Surprisingly, in Fig. 4(d), we find that the band asymmetry can change from positive to negative as the thickness reduces. All these observations suggest that the e-h asymmetry is more than just a small correction and allows a wide tuning range with the quantum confinement effect.

In conclusion, we present a magneto-IR spectroscopy study of the SSs in Sb$_2$Te$_3$ thin films grown by MBE with interface engineering. Due to  their ultralow carrier densities, these samples showed clear signatures of SSs related LL transitions, providing an ideal platform for the study of SSs, especially in magneto-optics. We find that the SSs are strongly modified by the e-h asymmetry from the bulk bands, giving rise to the shift of the DP and a band asymmetry between the conduction and valence SSs. The thickness reduction enhances the hybridization gap and also the band asymmetry in the SSs. We show that the correct modeling of the experimental results requires a self-consistent consideration of the bulk Hamiltonian and the nascent SSs. Further analysis reveals that the hybridization gap is strongly suppressed with e-h asymmetry, and the sign of band asymmetry can even be reversed with thicknesses. Our results strongly suggest the importance of e-h asymmetry and its interplay with thickness when considering the SS in the thin film limit, which provides important ingredients for future TI device applications.

We acknowledge the helpful discussion with Charles Kane. The magneto-IR measurements were performed at the National High Magnetic Field Laboratory (NHMFL), which is supported by the National Science Foundation (NSF) Cooperative agreement no. DMR-1644779 and the State of Florida. Y.J. acknowledges the support of the Jack E. Crow Postdoctoral Fellowship. L.W. acknowledges support by the Army Research Office under grant W911NF1910342 and a seed grant from NSF MRSEC at Penn under the grant DMR1720530. D.S. Y.J. and Z.J. acknowledge support under grant no. DE-FG02-07ER46451. Z.J. and L.W. acknowledge support from the NHMFL Visiting Scientist Program. M.M.A. and W.-K.T acknowledge support by the DOE Early Career Award no.DE-SC0019326. M.S. and S.O. are supported by the Gordon and Betty Moore Foundation’s EPiQS Initiative under grant no. GBMF4418 and the NSF under grant no. EFMA-1542798.

\end{document}


\title{Supporting Information: Electron-Hole Asymmetry of Surface States in Topological Insulator Sb$_2$Te$_3$ Thin Films Revealed by Magneto-Infrared Spectroscopy}
\author{Y. Jiang}
\affiliation{National High Magnetic Field Laboratory, Tallahassee, Florida 32310, USA}
\email{yjiang@magnet.fsu.edu, liangwu@sas.upenn.edu}
\author{M. M. Asmar}
\affiliation{Department of Physics and Astronomy, The University of Alabama, Tuscaloosa, Alabama 35487, USA}
\author{X. Y. Han}
\affiliation{Department of Physics and Astronomy, University of Pennsylvania, Philadelphia, Pennsylvania 19104, USA}
\author{M. Ozerov}
\affiliation{National High Magnetic Field Laboratory, Tallahassee, Florida 32310, USA}
\author{D. Smirnov}
\affiliation{National High Magnetic Field Laboratory, Tallahassee, Florida 32310, USA}
\author{M. Salehi}
\affiliation{Department of Material Science and Engineering, Rutgers, The State University of New Jersey,Piscataway, New Jersey 08854, USA}
\author{S. Oh}
\affiliation{Department of Physics and Astronomy, Rutgers, The State University of New Jersey,
Piscataway, New Jersey 08854, USA}
\author{Z. Jiang}
\affiliation{School of Physics, Georgia Institute of Technology, Atlanta, Georgia 30332, USA}
\author{W.-K. Tse}
\affiliation{Department of Physics and Astronomy, The University of Alabama, Tuscaloosa, Alabama 35487, USA}
\author{L. Wu}
\affiliation{Department of Physics and Astronomy, University of Pennsylvania, Philadelphia, Pennsylvania 19104, USA}
\date{\today}

\maketitle
\section{Sample structures and low field transport data}
\begin{figure*}[h!]
\includegraphics[width=16cm] {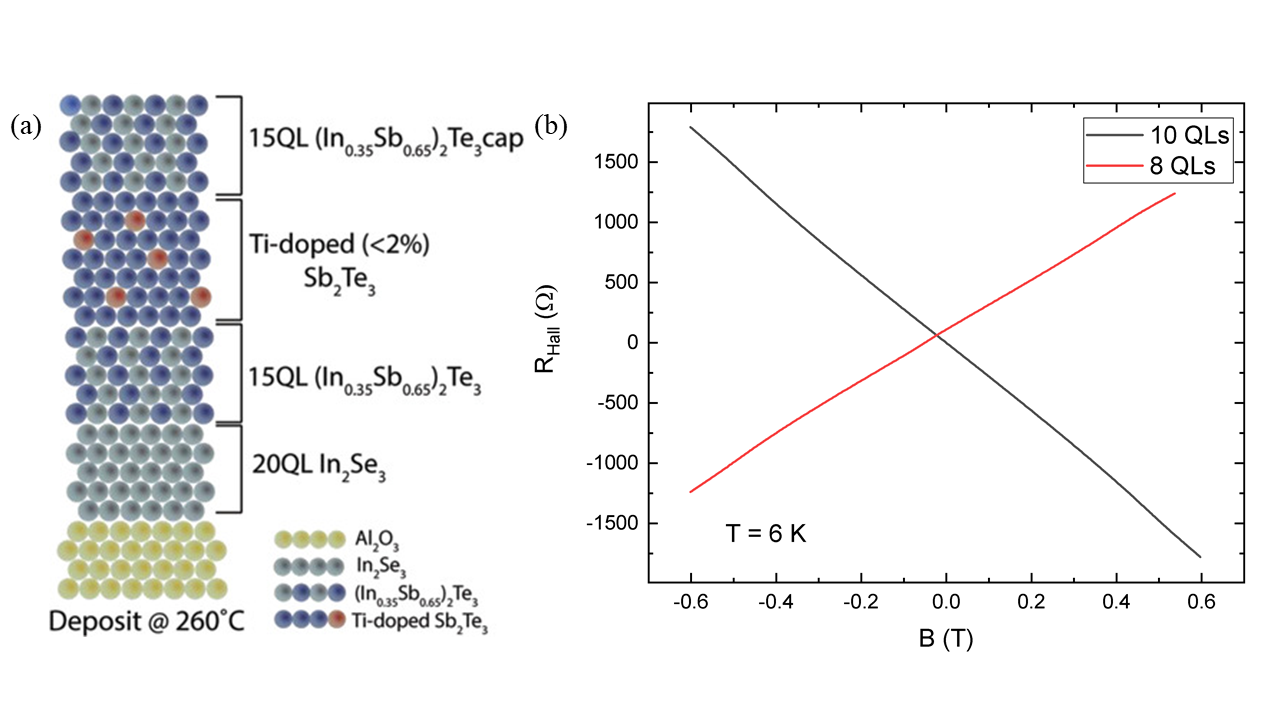}
\caption{(color online) (a) A schematic drawing of the sample structure. The samples studied here are from the same growth in Ref.\cite{Oh_AM}, and their thicknesses are also characterized therein. A 20-QL In$_2$Se$_3$ /15-QL (Sb$_0.65$In$_0.35$)$_2$Te$_3$ layer was first grown on a sapphire (Al$_2$O$_3$) substrate as a buffer layer to match the lattice constant of Sb$_2$Te$_3$. This buffer layer helps to reduce the defect density in the Sb$_2$Te$_3$ layer and allows tuning the carriers from p-type to n-type through titanium doping. Then, the Sb$_2$Te$_3$ layer is deposited and capped \textit{in situ} with another 15-QL (Sb$_0.65$In$_0.35$)$_2$Te$_3$ layer to protect it from aging and maintain a symmetric environment between its top and bottom surfaces. (b) Carrier concentrations in the studied samples. The carrier densities in the samples are characterized with magneto-transport at low temperature with the Van der Pauw geometry. The carrier types are different in these two samples due to different amounts of Titanium dopants. From the sign of the slope in the $R_{Hall}$ vs $B$ plot, the 8-QL sample is determined to be p-doped, and the 10-QL sample is n-doped. The extracted carrier densities from the slope of the data give $n_e= 2.2 \times 10^{11}$ cm$^{-2}$ and $n_h=3.1 \times 10^{11}$ cm$^{-2}$ for the 10-QL and 8-QL samples, respectively. }
\end{figure*}

\newpage
\section{Additional data on the 8-QL sample}
\begin{figure*}[h!]
\includegraphics[width=16cm] {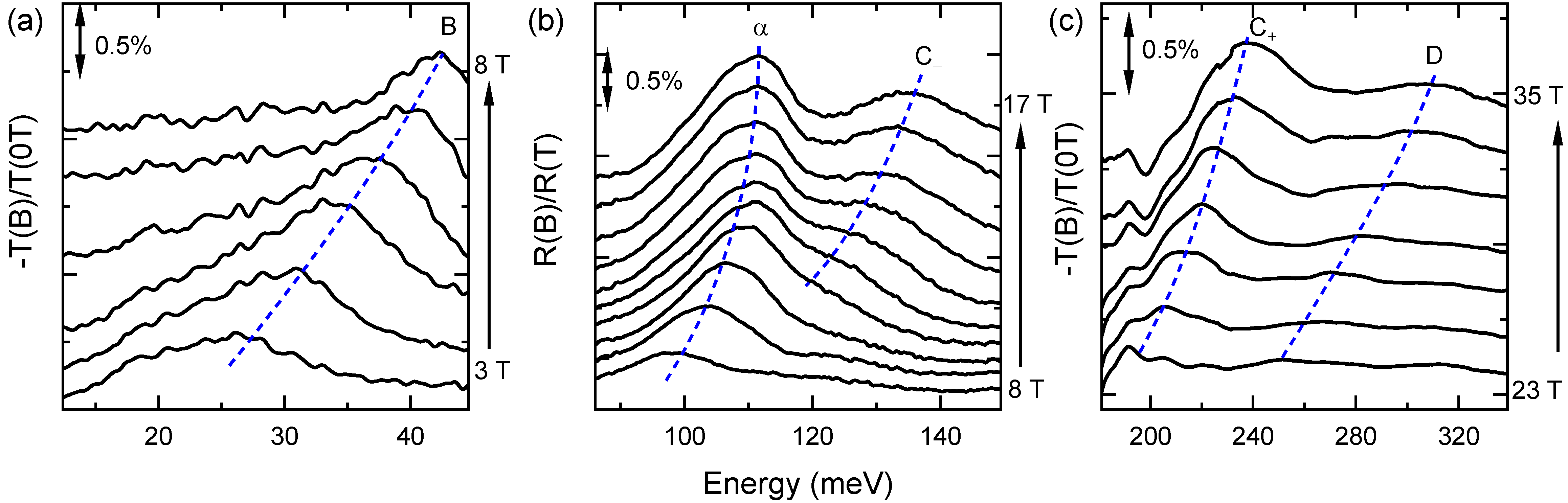}
\caption{(color online) Normalized magneto-infrared spectra of the 8-QL sample for (a)(c) transmission and (b) reflection measurements outside and inside the substrate Reststrahlen band, respectively. All the modes are labeled with a Latin or Greek letter. The blue dash lines are guides to the eye showing the mode evolution in magnetic field. Compared to the 10-QL sample, the A mode is not observable here, probably due to the broad linewidth. All spectra are shifted for clarity.}
\end{figure*}

\section{Symmetrization and anti-symmetrization of the surface state dispersion}
The conduction surface state (CSS) and valence surface state (VSS) can be written generally as \cite{eh_Theory1}:
\begin{align*}
    E_{CSS}=E_0+Ck^2+\sqrt{A^2k^2+(M-Bk^2)^2},\\
    E_{VSS}=E_0+Ck^2-\sqrt{A^2k^2+(M-Bk^2)^2}.
\end{align*}
The first term is the overall offset $E_0$, and the second term is the band asymmetry term described by $C$. The last term is the dominant component of the SS, which is described by the linear band parameter $A$, the Dirac mass $M$, and band inversion parameters $B$ and exhibits a electron-hole (e-h) symmetry. We can arrive at the e-h symmetry part $E_{sym}$ and e-h asymmetry part $E_{asy}$ by
\begin{align*}
    &E_{sym}=(E_{CSS}-E_{VSS})/2=\sqrt{A^2k^2+(M-Bk^2)^2},\\
    &E_{asy}=(E_{CSS}+E_{VSS})/2=Ck^2+E_0.
\end{align*}
From here, it is evident to see that the anti-symmetrized part leads to the band asymmetry term $Ck^2$ and an offset $E_0$, and the symmetrized part to the dominant dispersion also with an offset from the Dirac mass $M$. For clear comparisons between the dispersions of different thicknesses, we shift all the $E_{Sym}$ and $E_{Asy}$ to the same origin in Fig. 4(c-d) in the main text. 

Finally, we note that when the gap $M$ is sufficiently large, we can Taylor expand $E_{sym}$:
\begin{align*}
    E_{sym}=\sqrt{A^2k^2+(M-Bk^2)^2}=M+\frac{(A^2-2M B)k^2}{2M}+\dots,
\end{align*}
 which explains the quadratic behavior in the ultrathin film limit.